\documentclass[twocolumn,showpacs,pre,english,showkeys,aps,unsortedaddress,superscriptaddress]{revtex4-1}
\usepackage{amsmath}
\usepackage{epsfig}
\usepackage{color}
\usepackage{epstopdf}
\usepackage{amssymb}
\usepackage{multirow}
\usepackage{hyperref}

\begin{document}

\title{Evidence for Supersymmetry in the Random-Field Ising Model at $D = 5$}

\author{Nikolaos G.~Fytas}
\affiliation{Applied Mathematics Research Centre, Coventry
University, Coventry CV1 5FB, United Kingdom}

\author{V\'{i}ctor Mart\'{i}n-Mayor}
\affiliation{Departamento de F\'isica
  T\'eorica I, Universidad Complutense, 28040 Madrid, Spain}
\affiliation{Instituto de Biocomputac\'ion y F\'isica de Sistemas
Complejos
  (BIFI), 50009 Zaragoza, Spain}

\author{Giorgio Parisi}
\affiliation{Dipartimento  di  Fisica,  Sapienza Universit\`{a} di
Roma,  P.le  Aldo  Moro  2, 00185  Rome, Italy and INFN, Sezione
di Roma I,  IPCF -- CNR,  P.le  A. Moro 2, 00185 Rome, Italy}

\author{Marco Picco}
\affiliation{Laboratoire de Physique Th\'{e}orique et Hautes Energies, UMR7589, Sorbonne Universit\'{e} et CNRS, 4 Place Jussieu,
75252 Paris Cedex 05, France}

\author{Nicolas Sourlas}
\affiliation{Laboratoire de Physique Th\'eorique de l'Ecole
Normale Sup\'erieure (Unit{\'e} Mixte de Recherche du CNRS et de l'Ecole Normale
  Sup\'erieure, associ\'ee \`a l'Universit\'e Pierre et Marie Curie, PARIS VI)
  24 rue Lhomond, 75231 Paris Cedex 05, France}

\date{\today}

\begin{abstract}
\noindent We provide a non-trivial test of supersymmetry in the
random-field Ising model at five spatial dimensions, by means of
extensive zero-temperature numerical simulations. Indeed,
supersymmetry relates correlation functions in a $D$-dimensional
disordered system with some other correlation functions in a $D-2$
clean system. We first show how to check these relationships in a
finite-size scaling calculation, and then perform a high-accuracy
test. While the supersymmetric predictions are satisfied even to our
high-accuracy at $D=5$, they fail to describe our results at $D=4$.
\end{abstract}

\pacs{05.50.+q,75.10.Nr,02.60.Pn,75.50.Lk}

\maketitle

{\it Introduction.}--- The suggestion~\cite{parisi:79c} that the
random-field Ising model (RFIM) at the critical
point~\cite{imry:75,nattermann:98,belanger:98} obeys supersymmetry
came as a major surprise in Theoretical Physics.  One of the
implications of supersymmetry is dimensional
reduction~\cite{aharony:76,young:77}: the critical exponents of a
disordered system at space dimension $D$ and those of a pure
(i.e. non-disordered) system at dimension $D-2$ coincide. Let us
remark that dimensional reduction is a consequence
of~\cite{parisi:79c,cardy:83}, but not necessarily equivalent to,
supersymmetry.

However, in spite of its power and elegance, it was soon clear that
the applicability of supersymmetry is problematic.  The original
argument~\cite{parisi:79c} was based on the study of the solutions of
the stochastic Landau-Ginsburg equations in the presence of a random
magnetic field. Unfortunately, the crucial assumption of uniqueness of
the solution of these equations~\cite{parisi:79c} (which holds at all
orders in perturbation theory), fails beyond perturbation theory. In
fact, it was immediately clear that in the RFIM the predicted
dimensional reduction is absent at low dimensions (but not for
branched polymers~\cite{parisi:81} where dimensional reduction has
been mathematically proven~\cite{brydges:03,imbrie:03,cardy:03}): the RFIM has a ferromagnetic phase at
$D=3$~\cite{imbrie:84,bricmont:87} while the $D=1$ pure Ising model
has no transition. Non-perturbative effects (e.g.  bound-states in
replica space~\cite{parisi:94,parisi:02,brezin:98,brezin:01}) are
obviously important in $D=3$. Yet, their relevance for $D>3$
(specially upon approaching the presumed upper critical dimension
$D_{\mathrm{u}}=6$) is unclear. If we consider the case of
$D=6-\epsilon$, different scenarios are possible, as listed below:
\begin{enumerate}
\item Nonperturbative effects could destroy supersymmetry at a finite
  order in the $\epsilon$ expansion or, even worse, at $D=6$.
\item Violations of supersymmetry might be exponentially small  $\sim\exp (-A/\epsilon)$  (see e.g. Refs.~\cite{parisi:92,dotsenko:07}; the computation of $A$ is still an unsolved problem).
\item Supersymmetry has been suggested to be exact but only for $D>D_{\mathrm{int}}\approx 5.1$~\cite{tissier:11,tissier:12,tarjus:13}. For $D<D_{\mathrm{int}}$ the supersymmetric fixed point becomes unstable with respect to non-supersymmetric perturbations.
\end{enumerate}

In order to discriminate among these three scenarios, we need accurate
simulations aimed to test some of the many predictions of supersymmetry. In
the last few years, the development of a powerful panoply of simulation and
statistical analysis methods~\cite{fytas:13,fytas:15b,fytas:16} set the basis
for a fresh revision of the problem. Great emphasis was made on the anomalous
dimensions $\eta$ and $\overline{\eta}$ related to the decay of the
connected and disconnected correlations functions, respectively [see
Eq.~\eqref{eq:anomalous}]. Supersymmetry predicts $\eta=\overline{\eta}$
(moreover, the $D$-dimensional RFIM $\eta=\overline{\eta}$ are predicted to be
equal to the anomalous dimension of the pure Ising model in dimension
$D-2$). Extensive numerical simulations at zero temperature showed that these
relations fail at $D = 3$~\cite{fytas:13} and $D = 4$~\cite{fytas:16}, but they
are valid with good accuracy at $D = 5$~\cite{fytas:17}.  These numerical
results suggest that supersymmetry may be really at play at $D=5$. We should
mention as well a recent work using conformal boostrap~\cite{hikami:18}, 
where it was found that dimensional reduction holds in the RFIM for $D\geq 5$.

The predictions of supersymmetry go further beyond those regarding the critical
exponents: they involve both finite volume effects and high-order correlations
functions. Here, we will show that several non-trivial supersymmetry
predictions hold at $D=5$ to a very high numerical accuracy. This is the first
direct confirmation that supersymmetry holds in the RFIM at high
dimensions. As a consistency check, we show that the same relations are
definitively not-satisfied at $D=4$.

{\it Simulation setup.} --- The Hamiltonian of the RFIM is
\begin{equation}
\label{H} {\mathcal H} = - J \sum_{<xy>} S_x S_y - \sum_{x} h_x S_x \;
,
\end{equation}
with the spins $S_x = \pm 1$ on a hypercubic lattice in $D$
dimensions with nearest-neighbor ferromagnetic interactions and $h_x$
independent random magnetic fields with zero mean and variance
$\sigma^2$. Given our previous universality
confirmations~\cite{fytas:18}, we have restricted ourselves to
normal-distributed $h_x$. We work directly at zero
temperature~\cite{auriac:85,ogielski:86,middleton:01,middleton:02,middleton:02b}
because the relevant fixed point of the model lies there~\cite{villain:84,bray:85,fisher:86}. The system has a
ferromagnetic phase at small $\sigma$, that, upon increasing the
disorder, becomes paramagnetic at the critical point
$\sigma_\mathrm{c}$. Here, we work directly at $\sigma_\mathrm{c}$,
namely at $6.02395\approx \sigma_{\rm c}(D=5)$~\cite{fytas:17} and
at $4.17749\approx \sigma_{\rm c}(D=4)$~\cite{fytas:16}.

We consider \emph{two} correlation functions, namely the
connected and disconnected propagators, $C^{\mathrm{(con)}}_{xy}$ and
$C^{\mathrm{(dis)}}_{xy}$:
\begin{equation}\label{eq:anomalous}
C^{\mathrm{(con)}}_{xy}\equiv\frac{\partial\overline{\langle
  S_x\rangle}}{\partial h_y} \,,\
 C^{\mathrm{(dis)}}_{xy}\equiv\
\overline{\langle S_x\rangle\langle S_y\rangle}\,,
\end{equation}
where the $\langle \cdots \rangle$ are thermal mean values as
computed for a given realization, a \emph{sample}, of the random
fields $\{h_x\}$. Over-line refers to the average over the
samples. 

For each of these two propagators, we scrutinize the second moment
correlation lengths~\cite{amit:05}, as adapted to our geometrical
setting. In particular, our chosen geometry is an elongated hypercube with
periodic boundary conditions and linear dimensions $L_x=L_y=L_z=L$ and
$L_t=L_u=R L$ (at $D=4$ we chose $L_x=L_y=L$ and $L_z=L_t=R L$) with
aspect ratio $R\geq 1$. In fact, the supersymmetric identities that we
will check in the critical region hold in the limit $R\to\infty$, which should be
taken \emph{before} the standard thermodynamic limit.

We simulated lattice sizes in the range $L = 4 - 14$  at $D = 5$
($L = 4 - 28$ at $D = 4$) and aspect ratios $1\leq R \leq 5$. Additional simulations for $R = 10$ and $L \leq 10$ were performed at both
5D and 4D for consistency reasons. For each pair of ($L$, $R$)-values we computed
ground states for $10^5$ disorder samples. Our simulations and analysis closely
follows the methodology outined in our previous works at $D = 3$ and $4$~\cite{fytas:13,fytas:16} (for full technical details see Ref.~\cite{fytas:15b}).

{\it Supersymmetric predictions.} --- Let us consider a point in the
5D lattice, $\mathbf{r}=(\mathbf{x},\mathbf{u})$ where
$\mathbf{x}=(x,y,z)$ refers to the first three cartesian coordinates,
while $\mathbf{u}=(t,u)$. In a similar vein, for the 4D case, we
split $\mathbf{r}=(x,y,z,t)=(\mathbf{x},\mathbf{u})$ as
$\mathbf{x}=(x,y)$ and $\mathbf{u}=(z,t)$. The supersymmetric predictions
(see~\cite{parisi:82,cardy:83,klein:84,cardy:85} and Appendix \ref{sect:FS-SUSY} for a more paused exposition)
are particularly simple for disconnected correlation functions:
\begin{equation}\label{eq:susy-1}
C^{\mathrm{(dis),D}}_{\mathbf{x}_1,\mathbf{u};\mathbf{x}_2,\mathbf{u}} = {\cal Z}
G^{\mathrm{Ising},D-2}_{\mathbf{x}_1;\mathbf{x}_2}\,,
\end{equation}
where $G$ is the pure Ising model correlator, and ${\cal Z}$ is a
position independent normalization constant that will play no role
(see below). Note that the left-hand side depends on both linear
dimensions, $L$ and $R L$, while the right-hand side depends only on
$L$. Therefore, we must carefully consider under which conditions
Eq.~\eqref{eq:susy-1} is expected to hold. In a more conventional
study, one would require an hierarchy of length scales $LR\gg L \gg \xi
\gg 1$ (recall that $\xi$ is the correlation length), while we demand
for the $D-2$ Euclidean distance $\Vert \mathbf{x}_1-\mathbf{x_2}
\Vert / \xi \sim 1$. We shall put under stress Eq.~\eqref{eq:susy-1}
by demanding it to hold as well in the finite-size scaling regime
\begin{equation}\label{eq:susy-2}
LR\gg L \sim \xi\gg 1\  ,\ \Vert \mathbf{x}_1-\mathbf{x_2}
\Vert / \xi \sim 1\,.
\end{equation}

These preliminaries lead us to consider a $D-2$ Fourier transform in the $D$-dimensional RFIM
\begin{equation}
\hat C^{\mathrm{(dis),D}}_{\mathbf{k}} =
\frac{1}{L^{D-2}} \sum_{\mathbf{x}_1,\mathbf{x}_2} \mathrm{e}^{\mathrm{i} {(\mathbf{x}_1-\mathbf{x}_2)\cdot\mathbf{k}}} \, \overline{\langle S_{\mathbf{x}_1,\mathbf{u}}\rangle\langle S_{\mathbf{x}_2,\mathbf{u}}\rangle}\,.
\end{equation}
Note that the $\mathbf{u}$-dependence vanishes due to the
disorder-average (hence we average over $\mathbf{u}$ in order to gain
statistics).  We then compute the second-moment correlation length
from the ratio of $\hat C^{\mathrm{(dis),D}}_{\mathbf{k}}$ at
$\mathbf{k}=\mathbf{0}$ and
$\mathbf{k}_\mathrm{min}=(2\pi/L,0,0)$~\cite{amit:05}
[$\mathbf{k}_\mathrm{min}=(2\pi/L,0)$ for $D=4$]. The important
observation is that, because the constant ${\cal Z}$ in the r.h.s. of
Eq.~\eqref{eq:susy-1} cancels when computing the ratio, the
dimensionless ratio $\xi^{(\mathrm{dis})}/L$ as computed in the
$D$-dimensional RFIM coincides with $\xi/L$ as computed in the $D-2$
Ising model. This equality holds if $\xi^{(\mathrm{dis})}/L$ is computed
precisely at the critical point $\sigma_\mathrm{c}$ and if the
thermodynamic limit is taken under conditions~\eqref{eq:susy-2}.

If we now consider the four-body disconnected correlation function,
supersymmetry predicts a relation analogous to Eq.~\eqref{eq:susy-1}
(the normalization in the r.h.s changes to ${\cal Z}^2$), so we may compute
as well a $(D-2)$-dimensional $U_4$ parameter,
\begin{equation}\label{eq:susy-3}
M_{\mathbf{u}} =\sum_{\mathbf x} S_{\mathbf{x},\mathbf{u}}\ ,\ U_4=\overline{\langle M_{\mathbf{u}}^4\rangle}/\overline{{\langle M_{\mathbf{u}}^2\rangle}}^2\,,
\end{equation}
that is predicted to coincide with that of the critical $D-2$ Ising model (under the same condition discussed above for
$\xi^{(\mathrm{dis})}/L$). Again, we improve our statistics by averaging both
$\overline{\langle M_{\mathbf{u}}^4\rangle}$ and $\overline{\langle
  M_{\mathbf{u}}^2\rangle}$ over $\mathbf{u}$.

We finally address the supersymmetric predictions for the connected
correlation function. It is convenient to consider the  correlation functions $K$ defined as
 \begin{equation}
K_{\mathbf{x}_1;\mathbf{x}_2} = \sum_{\mathbf{u}}  C^{\mathrm{(con)}}_{\mathbf{x}_1,\mathbf{0};\mathbf{x}_2,\mathbf{u}}\,.
\end{equation}
The Ward identity for supersymmetry~\cite{parisi:82} implies,
see Appendix~\ref{sect:Ward}, that the second-moment correlation length
$\xi^{\mathrm{(con)}}_{\sigma-\eta}$ computed from $K$~\footnote{We introduce the Fourier transform in $(D-2)$ dimensions, $\hat K(\mathbf{k})= 
\sum_{\mathbf{x}_1,\mathbf{x}_2} \mathrm{e}^{\mathrm{i} {(\mathbf{x}_1-\mathbf{x}_2)\cdot\mathbf{k}}} K_{\mathbf{x}_1,\mathbf{x}_2}/L^{D-2}$ and compute $\xi^{\mathrm{(con)}}_{\sigma-\eta}=[(\hat K(\mathbf{0})-\hat K(\mathbf{k}_\mathrm{min})/\hat K(\mathbf{k}_\mathrm{min})]^{1/2}/(2\sin\pi/L)$. For an extended discussion of the second-moment correlation length see, for instance, Ref.~\cite{amit:05}.}   is equal to the
disconnected correlations length. This prediction
$\xi^{\mathrm{(con)}}_{\sigma-\eta} =\xi^{(\mathrm{dis})}$ does not make
direct reference to dimensional reduction.
  
{\it Results.} --- Let us start by recalling in
Table~\ref{tab:pure-Ising} the $(D-2)=2,3$ universal quantities from
the pure Ising model that we aim to recover from the $D$
dimensional RFIM. We shall need as well the value of the leading
corrections to scaling exponent $\omega$); the analysis we present is done using the exponent $\omega$ given by dimensional reduction, which is not far from the one computed in the large-scale simulations at $D=5$~\cite{fytas:17}. 

\begin{table}[b]
\caption{\label{tab:pure-Ising} Universal quantities as computed in the
pure Ising model at two and three spatial dimensions. The somewhat
controversial situation with the corrections to
scaling exponent $\omega$ in two dimensions is discussed in Appendix~\ref{sect:omega-Ising-2D}.}
\begin{ruledtabular}
\begin{tabular}{clll}
$D-2$ & \hphantom{aaaaa} $\xi/L$ & \hphantom{aaaaa}$U_4$ & 
\hphantom{aaaaa}$\omega$ \\
2 & 0.9050488\ldots\,\cite{salas:00} & 1.16793\ldots\,\cite{salas:00} & 1.75\\
3  &  0.6431(1)\,\cite{hasenbusch:10} & 
1.6036(1)\,\cite{hasenbusch:10}  &  0.82966(9)\,\cite{kos:16}
\end{tabular}
\end{ruledtabular}
\end{table}

\begin{figure}
\centering
\includegraphics[width=\columnwidth,trim=70 60 100 70]{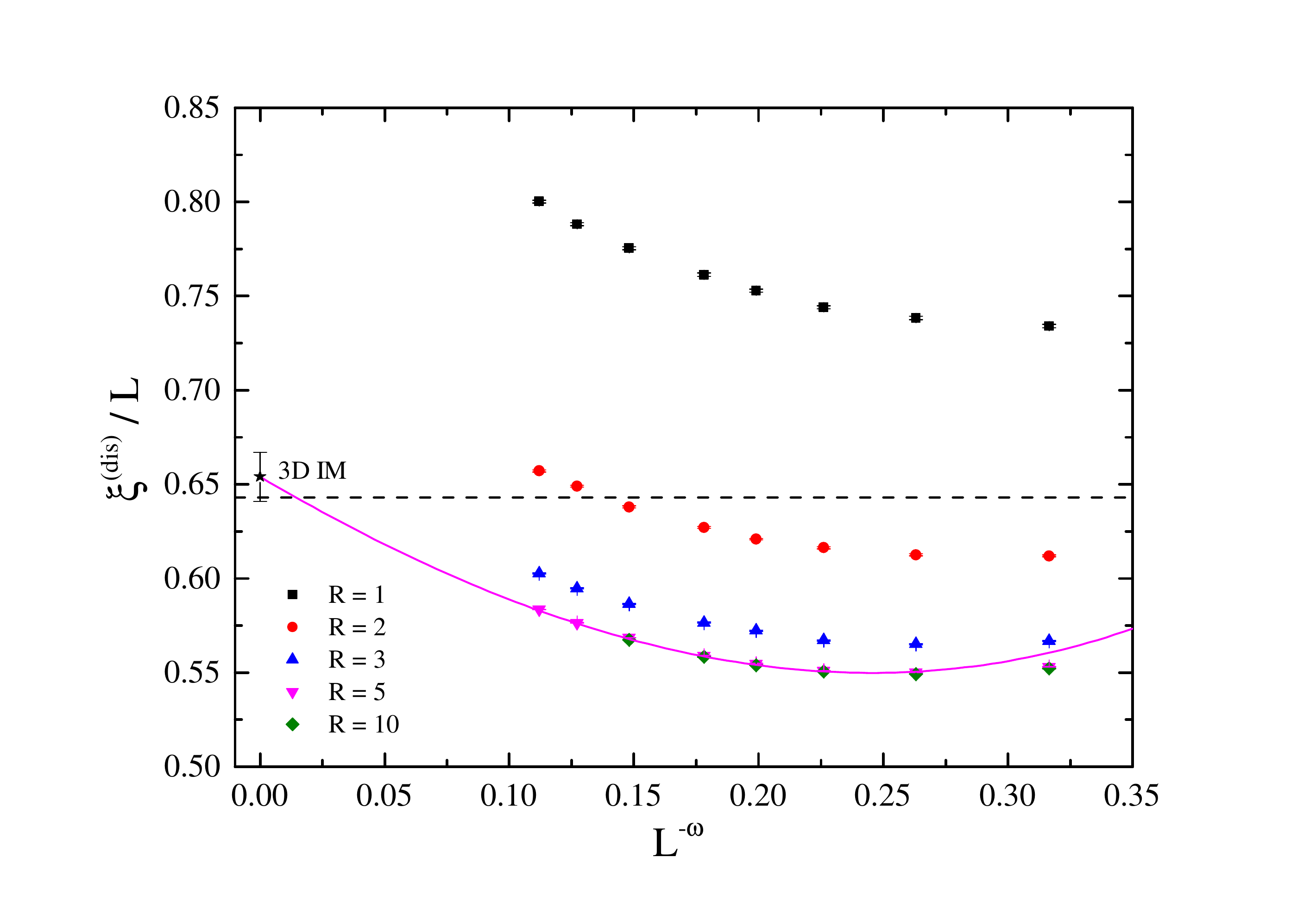}
\caption{\label{fig:xi_dis} $\xi^{\mathrm{(dis)}}(L,R)/L$ vs.
  $L^{-\omega}$ for various $R$ values, as computed in the $D=5$ RFIM. The value  of the corrections to scaling exponent $\omega$ corresponds to
  the pure Ising model in three spatial dimensions, see
  Table~\ref{tab:pure-Ising} (the value from Ref.~\cite{kos:16} is so
  accurate that we took their central value as numerically exact).  The
  dashed horizontal line corresponds to the value for $\xi/L$, also shown in
  Table~\ref{tab:pure-Ising}. The continuous line is a fit to our
  $R=5$ data (see text for details). The extrapolation to $L=\infty$
  obtained from the fit is compatible with the pure Ising model value,
  as predicted by supersymmetry.}
\end{figure}

First, we consider the dimensionless ratio $\xi^{\mathrm{(dis)}}(L,R)/L$ in
Fig.~\ref{fig:xi_dis}. Our first task, recall Eq.~\eqref{eq:susy-2}, is to
extract the large-$R$ limit. The good news is that we expect this limit to be
reached exponentially in $R$ and uniformly in $L$~\footnote{Because we shall
  be taking the limit of large $R$ at fixed $L$, the gap in the transfer
  matrix scales as $1/L$. Therefore, correlation functions along the $t$ and
  $u$ axes ($z$ and $t$ axes at $D=4$), decay exponentially in $R$, for any
  $L$.}. In fact, the comparison of our numerical results for $R=5$ and $10$
suggests that (within our statistical accuracy) $R=5$ is large
enough. Therefore, we focus the analysis on $R=5$, where we reach our largest
$L$ value, namely $L=14$. As it is clear from Fig.~\ref{fig:xi_dis}, our data
are accurate enough to resolve corrections to scaling. Furthermore, the
non-monotonic $L$-evolution of $\xi^{\mathrm{(dis)}}(L,R=5)/L$ implies that
sub-leading corrections cannot be neglected. Hence, we have attempted to
represent these sub-leading corrections in an effective way by means of a fit
to a polynomial in $L^{-\omega}$. We have included in the fit only data with
$L\geq L_\mathrm{min}$. We have attempted to keep both $L_\mathrm{min}$ and
the order of the polynomial as low as possible. We find a fair fit
($\chi^2/\mathrm{dof}=3.24/2$, $p$-value=20\%) with a cubic polynomial and
$L_\mathrm{min}=6$. The corresponding extrapolation to $L=\infty$ is
\begin{equation}
 \lim_{L\to\infty} \, \left(\lim_{R\to\infty}\  \frac{\xi^{\mathrm{(dis)}}(L,R)}{L}\right)=0.654(13)\,,
\end{equation}
which is statistically compatible to the three-dimensional result in
Table~\ref{tab:pure-Ising}. Hence, our first check of supersymmetry has
been passed. The strength of this check is quantified by our  2\% accuracy.

\begin{figure}
\centering
\includegraphics[width=\columnwidth,trim=70 60 100 70]{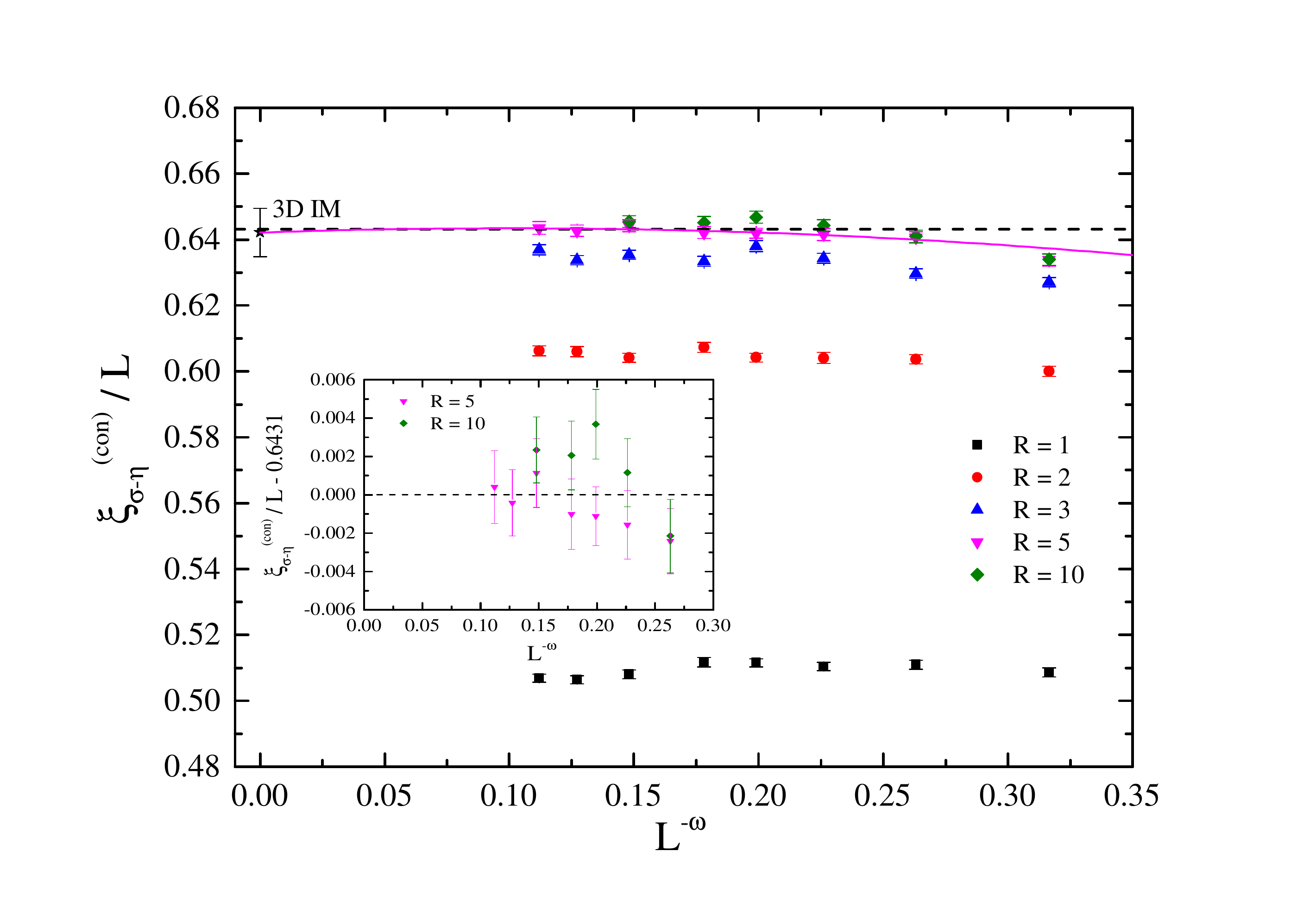}
\caption{\label{fig:xi_seta_con} As in Fig.~\ref{fig:xi_dis}, but for the
  $\xi^{\mathrm{(con)}}_{\sigma-\eta}(L,R)/L$ data, as computed in the $D = 5$            RFIM. The agreement of the $L=\infty$ extrapolation with the value of
  $\xi/L$ from the pure Ising model is a direct confirmation of the
  supersymmetric Ward identity, see Appendix~\ref{sect:Ward}. {\bf Inset:} Zoom of
  main panel data corresponding to $R=5$, $10$, and $L > 4$. For the sake of
  clarity, in the vertical axis, we have subtracted the value of the pure Ising model (see also Table~\ref{tab:pure-Ising}).}
\end{figure}
The analysis of $\xi^{\mathrm{(con)}}_{\sigma-\eta}(L,R)/L$, see Fig.~\ref{fig:xi_seta_con} is carried out along the same lines. We find a good fit
($\chi^2/\mathrm{dof}=0.63/3$, $p$-value=89\%) with a quadratic polynomial
in $L^{-\omega}$ and $L_\mathrm{min}=6$. The corresponding extrapolation to $L=\infty$ is 
\begin{equation}
 \lim_{L\to\infty} \,\left(\lim_{R\to\infty}\  \frac{\xi^{\mathrm{(con)}}_{\sigma-\eta}(L,R)}{L}\right)=0.642(7)\,.
\end{equation}
It follows that we have checked supersymmetry to a 1\% accuracy.

\begin{figure}
\centering
\includegraphics[width=\columnwidth,trim=70 60 100 70]{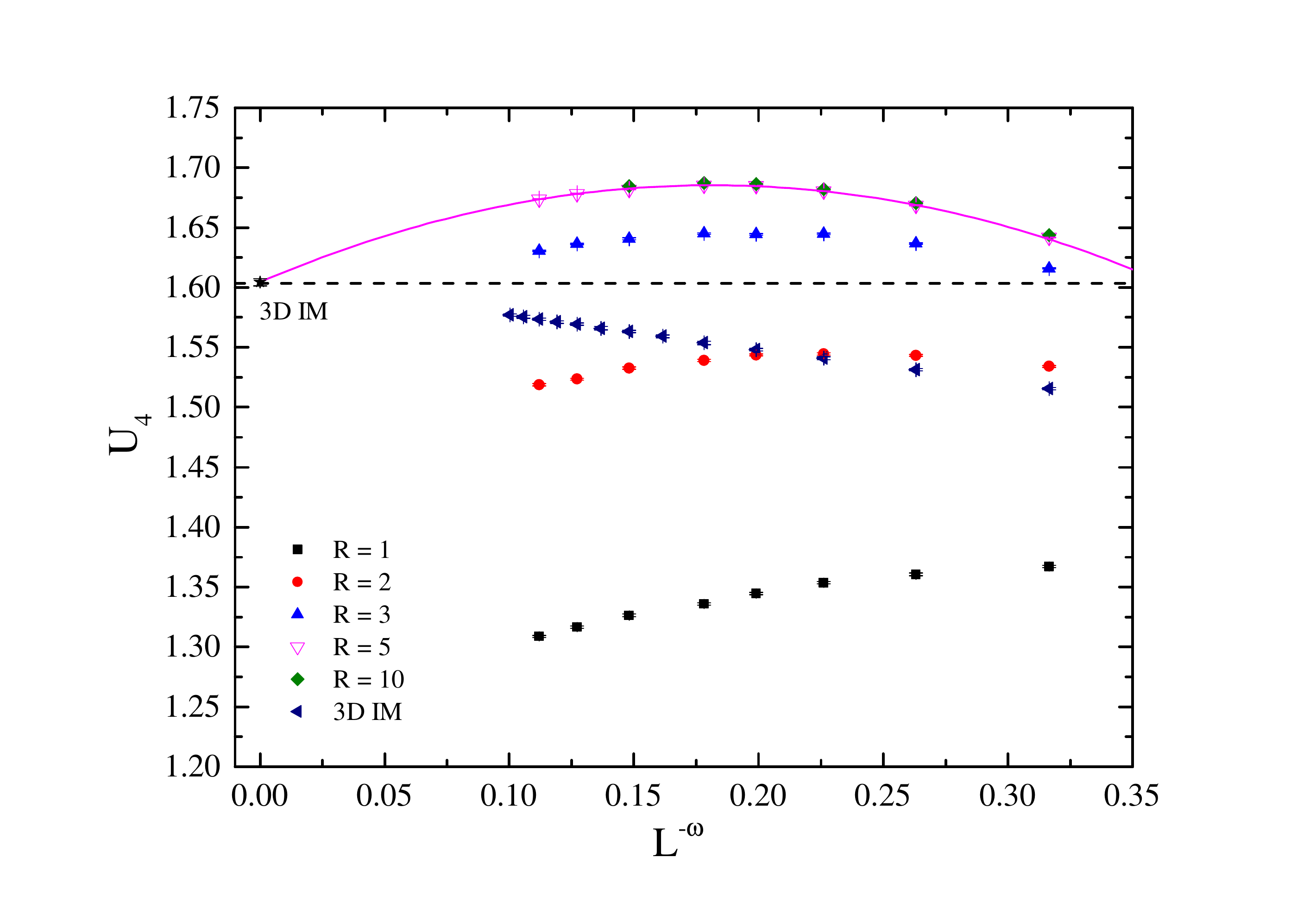}
\caption{\label{fig:U4} As in Fig.~\ref{fig:xi_dis}, but for the $U_4(L,R)$ data, as computed in the $D = 5$ RFIM. For comparison,
  we also show data for the pure Ising model in three spatial
  dimensions.  Corrections to scaling in the pure model are of similar
  size (but opposite sign) to those of the large $R$ limit for the
  $RFIM$ at $D = 5$.}
\end{figure}
Our $U_4(L,R)$ data, see Fig.~\ref{fig:U4}, can be analyzed in a similar
vein. We find a fair fit ($\chi^2/\mathrm{dof}=6.85/4$, $p$-value=14\%) with a quadratic polynomial in $L^{-\omega}$ and $L_\mathrm{min}=5$. The corresponding extrapolation to $L=\infty$ is 
\begin{equation}
 \lim_{L\to\infty} \,\left(\lim_{R\to\infty}\  U_4(L,R)\right) =1.604(3)\,,
\end{equation}
again compatible with the three-dimensional pure Ising model value
(Table~\ref{tab:pure-Ising}). Supersymmetry is checked to the 0.2\%
level, this time.

\begin{figure}[t]
\centering
\includegraphics[width=\columnwidth,trim=0 0 0 0]{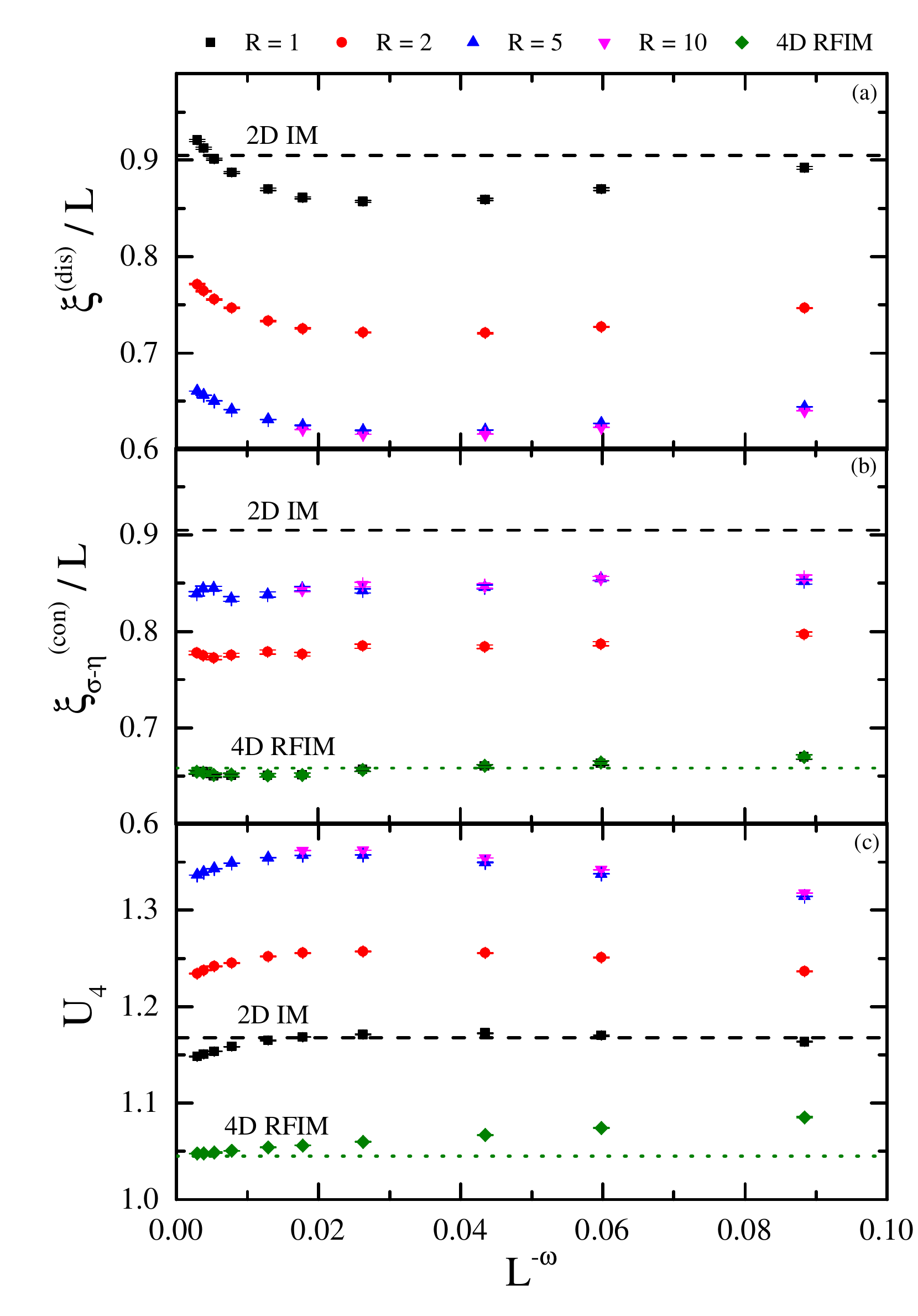}
\caption{\label{fig:4D} Dimensionless quantities
  $\xi^{\mathrm{(dis)}}(L,R)/L$ ({\bf a}),
  $\xi^{\mathrm{(con)}}_{\sigma-\eta}(L,R)/L$ ({\bf b}) and
  $U_4(L,R)$ ({\bf c}) vs. $L^{-\omega}$ as computed in the $D = 4$ RFIM.
  We set $\omega=1.75$ from Table~\ref{tab:pure-Ising}. We
  show the corresponding universal values for the 2D pure
  Ising model (black dashed lines).  Note that for $R=1$ there are
  two natural ways of computing $U_4$. One way (black squares) is
  averaging over a co-dimension two manifold [this is the natural way
  for a supersymmetry check, recall Eq.~\eqref{eq:susy-3}]. The
  other way, which is the natural one when studying the $D=4$ RFIM
  \emph{per se}, is averaging over the full four-dimensional lattice
  (green diamonds).  Clearly, the two choices differ, both at finite
  $L$ and in the large-$L$ limit. Instead, for
  $\xi^{\mathrm{(con)}}_{\sigma-\eta}(L,R)/L$ these two kinds of
  spatial-averaging coincide by construction. The horizontal green
  dotted lines are the large-$L$ limit, as obtained for the $D=4$
  RFIM~\cite{fytas:16}.}
\end{figure}

Finally, as a comparison, we show our data for the 4D RFIM Ising model in
Fig~\ref{fig:4D}. Even after carrying out the double limit $L\rightarrow \infty$
and $R\rightarrow \infty$, all three dimensionless quantities differ from their values in the 2D pure Ising ferromagnet. Although this is hardly a surprise
(recall, for instance, exponents $\eta$ and $\overline{\eta}$~\cite{fytas:16}), the discrepancy is at least at the 10\% level.

{\it Conclusions.}--- The finding of supersymmetry and dimensional reduction in
the RFIM is, arguably, one of the most surprising results in Theoretical
Physics. Here, thanks to state-of-the-art numerical techniques, we have
carried out a precision test of supersymmetry. Although supersymmetry is clearly
broken at $D=4$, the $D=5$ RFIM is supersymmetric with good accuracy. Hence,
the Scenario 1 in the Introduction is plainly discarded.

The only remaining contenders are Scenarios 2 and 3. Exponent $\omega$
might help to settle the question. In the $\epsilon$ expansion
($\epsilon=6-D$) we find at least two exponents:
$\omega_{\mathrm{DR}}=\epsilon+{\cal O}(\epsilon^2)$ (obtained through
dimensional reduction) and $\omega_{\mathrm{NS}}=2+{\cal
  O}(\epsilon^2)$ (due to irrelevant non-supersymmetric
operators). The large value of $\omega$ found here and in
Ref.~\cite{fytas:17} (the values for $\omega(D)$ are
in Appendix~\ref{sect:omega-RFIM}), agrees with dimensional reduction and favors
Scenario 2. Indeed, in Scenario 3 supersymmetry is broken only for
space dimension $D <D_{\mathrm{int}}$, suggesting a much smaller value
$\omega(D=5)\sim D_{\mathrm{int}}-D\approx 0.1$. However, further
studies are needed to resolve this delicate issue.

\begin{acknowledgments}
We acknowledge partial financial support from Ministerio de
Econom\'ia, Industria y Competitividad (MINECO, Spain) through Grant
No. FIS2015-65078-C2, and from the European Research Council (ERC)
under the European Union's Horizon 2020 research and innovation
program (Grant No. 694925). N.~G.~F. and M.~P. were supported by a
Royal Society International Exchanges Scheme 2016/R1.
\end{acknowledgments}

\newpage

\appendix

\section{Finite volume supersymmetry}
\label{sect:FS-SUSY}

In the case of RFIM in the Landau-Ginsburg form, it is well known that we can
neglect the thermal fluctuations near the critical temperature and the model
becomes equivalent to a stochastic differential equation.  Under the {\sl
  approximation} of uniqueness of the solution, we arrive to a supersymmetric
field theory. In this theory we can define the superfield $\Phi( X)$ as
function of the superposition $ X=x\oplus \theta$, 
\begin{equation}\label{eqSI:susy-1}
\Phi(X)=\phi(x)+\bar\theta \psi(x)+\bar\psi(x) \theta + \bar\theta \theta \lambda(x)\,,
\end{equation}
where $\theta$ is a complex anticommuting quantity, $\phi(x)$ is the
original field and $\psi(x)$ and $\lambda(x)$ are auxiliary fields,
whose correlations functions are related to the response functions.
For instance,  in the supersymmetric formulation
the connected propagator $C^{\mathrm{(con)}}_{xy}$ corresponds to the
propagator of the fermionic field
$\langle \bar\psi(x) \psi(y)\rangle$, while the disconnected propagator
$C^{\mathrm{(dis)}}_{xy}$ corresponds to the propagator for the bosonic field
$\langle \phi(x) \phi(y)\rangle$. 

In the infinite volume limit, the theory is invariant  under
the supergroup $O(D|2)$ which implies that the correlation functions
are functions of the superdistances. In particular, the correlation
function $\langle\Phi(X) \Phi(Y)\rangle$ is a function of
\begin{equation}\label{eqSI:susy-2}
(X-Y)^2=r^2+(\theta_x -\theta_y)(\overline\theta_x -\overline\theta_y) \; ,
\end{equation}
where $r^2$ is the (squared) Euclidean distance between points $x$ and $y$ in
the $D$-dimensional space:
\begin{equation}\label{eqSI:susy-2b}
\langle\Phi(x)
\Phi(y)\rangle = \langle\phi(x) \phi(y)\rangle + \bar\theta \theta
\langle\bar\psi(x) \psi(y)\rangle= F(Z)
\end{equation}
 where $Z=(X-Y)^2$.  By
Taylor expanding both sides of Eq.~\eqref{eqSI:susy-2b} in powers of $\bar\theta \theta$ we conclude that
\begin{equation}\label{eqSI:susy-2c}
F(Z) = F(r^2) +
\bar\theta \theta F'(r^2)\,,
\end{equation}
 because all higher powers of $ \bar\theta \theta $ vanish.
We readily obtain the Ward identity~\cite{parisi:82}
\begin{equation}\label{eqSI:susy-2d}
\langle\bar\psi(x)
\psi(y) \rangle= - \frac{\mathrm{d}\langle\phi(x) \phi(y)\rangle}{\mathrm{d}r^2}\,.
\end{equation}
We note that 
Eq.~\eqref{eqSI:susy-2d} implies for the RFIM in a infinite lattice that
\begin{equation}\label{eqSI:susy-4}
C^{\mathrm{(con)}}_{r}= -{\cal Z}_2\frac{\mathrm{d}}{\mathrm{d} r^2} C^{\mathrm{(dis)}}_{r}\,,
\end{equation}
where large $r$ and $\xi$ are assumed ($\xi$ is the correlation
length), so that $D$-dimensional rotational invariance is restored,
and ${\cal Z}_2$ is a position-independent (therefore, irrelevant for
us) constant\footnote{When combined with the long distance decay of
  the propagators at the critical point, $C^{\mathrm{(con)}}_{r}\sim
  1/r^{D-2+\eta}$ and $C^{\mathrm{(dis)}}_{r}\sim 1/r^{D-4+\bar\eta}$,
  the Ward identity~\eqref{eqSI:susy-4} tells us that
  $\eta=\bar\eta$.}.  These relations
(\ref{eqSI:susy-2b}-\ref{eqSI:susy-4}) lead to a bunch of Ward
identities among various correlation functions.  One also finds that
the probability distribution of the $\phi$ field on a $d\equiv
D-2$-dimensional hyperplane is the same of the dimensional reduced
theory.

However, in a finite volume rotational invariance is broken so that
supersymmetry and dimensional reduction are lost.  Fortunately close
examination of the argument shows that we do not need the full
$O(D|2)$ supersymmetry, but the $O(2|2)$ supersymmetry is enough in
order to have dimensional reduction. In order to recover the $O(2|2)$
supersymmetry, the system size needs to be infinite only in the
remaining two dimensions.

 Our choice (see main text) is to stay in a system of linear size $L$ in $d$
 directions and of size $LR$ in two directions. At the end we need to consider
 the limit $R\to\infty$ in order to have supersymmetry and dimensional
 reduction.  Let us write the $D$ dimensional coordinates $\mathbf{r}$ as
 $(\mathbf{x},\mathbf{u})$, where $x$ is $d$-dimensional and $u$ is two
 dimensional. We can write
 \begin{equation}\label{eqSI:susy-3}
 X=\mathbf{r}\oplus \theta=\mathbf{x}\oplus \mathbf{u}\oplus \theta \,.
\end{equation}
 The $O(2|2)$ supersymmetry acts on the two-dimensional subspace,
 labeled by coordinates $\mathbf{u}\oplus \theta $, that becomes
 infinite in the $R\to\infty$ limit.  Dimensional reduction gives
 informations only on the probability distribution on fields on the
 hyperplanes at fixed $\mathbf{u}$ that have volume $L^d$.

Supersymmetry does not give us information on the behaviour of the
correlations function of fields whose $\mathbf{u}$ is different, unless we
stay at distances much smaller than $L$, where $2+d$ rotational invariance is
recovered. It connects however responce functions at different $\mathbf{u}$
with the correlations functions at fixed $\mathbf{u}$, as we shall see below.

\section{The Ward Identity and its consequences}
\label{sect:Ward}

As explained above (see also main text), we shall be considering points in the
five-dimensional lattice, $\mathbf{r}=(\mathbf{x},\mathbf{u})$ where
$\mathbf{x}=(x,y,z)$ refers to the first three cartesian coordinates, while
$\mathbf{u}=(t,u)$. In a similar vein, for the $D=4$ case, we split
$\mathbf{r}=(x,y,z,t)=(\mathbf{x},\mathbf{u})$ as $\mathbf{x}=(x,y)$ and
$\mathbf{u}=(z,t)$. The (squared) Euclidean distance between two points in the
$D$ dimensional lattice will be named $r^2=\mathbf{x}^2+\rho^2$ (in $D=5$,
$\rho^2=t^2+u^2$, while in $D=4$ we have $\rho^2=z^2+t^2$).

In the finite $L$ case we only have a $O(2|2)$
supersymmetry. Therefore, instead of the Ward identities corresponding
to $O(D|2)$, see Eqs.~(\ref{eqSI:susy-2},\ref{eqSI:susy-4}), the
Bosonic and Fermionic propagators are now related through a $O(2|2)$ Ward
identity that tells us that
\begin{equation}\label{eqSI:susy-5}
C^{\mathrm{(con)}}_{\mathbf{x},\mathbf{u}}= -{\cal Z}_2\frac{\mathrm{d}}{\mathrm{d} \rho^2} C^{\mathrm{(dis)}}_{\mathbf{x},\mathbf{u}}\,.
\end{equation}
In our geometry, we only have the full $D$-dimensional rotational symmetry
for $\mathbf{x}^2\ll L^2$. Instead, in the limit of a large aspect ratio, $R\to\infty$,
we have two-dimensional rotational symmetry (for the $\mathbf{u}$ variables)
for any ${\mathbf x}$. Thus, we expect the two correlation functions
$C^{\mathrm{(dis)}}_{\mathbf{x},\mathbf{u}}$ and $C^{\mathrm{(con)}}_{\mathbf{x},\mathbf{u}}$
to be functions of
\begin{equation}\label{eqSI:susy-6}
  g(\mathbf{x})+\rho^2\,,
\end{equation}
where $g(\mathbf{x})$ is some function of the $d$-dimensional coordinates that
reduces to the $d$-dimensional Euclidean distance $\mathbf{x}^2$ in the limit $\mathbf{x}^2\ll
L^2$ [a simple possibility in $D=5$ would be $g(\mathbf{x})=L^2 \pi^{-2} (\sin^2 \pi
  x/L + \sin^2 \pi y/L + \sin^2 \pi z/L)$].

Let us now consider the $\mathbf{u}$-averaged correlation function
\begin{equation}
K_{\mathbf{x}_1;\mathbf{x}_2} = \sum_{\mathbf{u}}  C^{\mathrm{(con)}}_{\mathbf{x}_1,\mathbf{0};\mathbf{x}_2,\mathbf{u}}\,,
\end{equation}

The $D=5$ reasoning goes as
follows (the $D=4$ case is analogous):
\begin{equation}
K_{\mathbf{x}_1;\mathbf{x}_2}\approx \iint_{-\infty}^{\infty}
\,\mathrm{d}t\,\mathrm{d}u\,
C^{\mathrm{(con)}}_{\mathbf{x}_1,0,0;\mathbf{x}_2,t,u}\,.\label{eqSI:susy-7a}
\end{equation}
We now introduce polar
  coordinates in the $(t,u)$ plane, $t=\rho\cos\varphi$ and
  $u=\rho\sin\varphi$:
\begin{equation}
K_{\mathbf{x}_1;\mathbf{x}_2}\approx \pi \int_0^{\infty}\mathrm{d}\rho^2\,
  C^{\mathrm{(con)}}_{\mathbf{x}_1,0,0;\mathbf{x}_2,\rho,0}\,.\label{eqSI:susy-7b}
\end{equation}
Our next step, will be using the Ward identity~\eqref{eqSI:susy-5}:
\begin{equation}
K_{\mathbf{x}_1;\mathbf{x}_2}\approx\pi {\cal Z}_2 \int_0^{\infty}\mathrm{d}\rho^2\,
  \left[-\frac{\mathrm{d}}{\mathrm{d} \rho^2}
    C^{\mathrm{(dis)}}_{\mathbf{x}_1,0,0;\mathbf{x}_2,\rho,0}\right]\,,\label{eqSI:susy-7c}
\end{equation}
and thus, we finally get
\begin{equation}
K_{\mathbf{x}_1;\mathbf{x}_2}\approx \pi {\cal Z}_2\,C^{\mathrm{(dis)}}_{\mathbf{x}_1,0,0;\mathbf{x}_2,0,0}\,.\label{eqSI:susy-7d}
\end{equation}
Note that, because we shall be taking the limit of large $R$ at fixed $L$, the
gap in the transfer matrix scales as $1/L$. Therefore, the correlation
function $C^{\mathrm{(dis)}}_{\mathbf{x}_1,0,0;\mathbf{x}_2,\rho,0}$ decays
exponentially in $\rho$ (for any $L$), so the convergence of the
two-dimensional integrals in Eqs.~\eqref{eqSI:susy-7a}--\eqref{eqSI:susy-7c}
poses no problems.

Hence, in the large-$R$ limit, the second-moment correlation length
$\xi^{\mathrm{(con)}}_{\sigma-\eta}$ is predicted to coincide with the one
obtained from the disconnected propagator. The prediction holds to a high
accuracy in the RFIM in $D=5$, but certainly not in $D=4$ (see Fig. 4 in the
main part).

Let us conclude this section by explainig our naming
$\xi^{\mathrm{(con)}}_{\sigma-\eta}$ to the correlation length extracted from
the $K$ propagator, which stems from the way it is computed. Indeed, the
Fluctuation-Dissipation relations for Gaussian random-fields~\cite{fytas:15b}
suggest a simple way to compute the $K_{\mathbf{x}_1;\mathbf{x}_2}$
propagator. Let
$$\sigma(\mathbf{x})=\sum_{\mathbf{u}} S_{\mathbf{x},\mathbf{u}}\,,\ \eta(\mathbf{x})=
\sum_{\mathbf{u}} h_{\mathbf{x},\mathbf{u}}\,,$$
then
$$K_{\mathbf{x}_1;\mathbf{x}_2} =\frac{1}{R^2L^2}\overline{\langle \sigma(\mathbf{x_1}) \eta(\mathbf{x_2})\rangle}\,.$$ Of course, $\mathbf{x_1}$ and $\mathbf{x_2}$
might be interchanged, so it is better to average over the two orderings.

\section{Exponent $\omega$ for the RFIM: the smoking gun?}
\label{sect:omega-RFIM}

\begin{figure}[h]
\centering
\includegraphics[width=\columnwidth,trim=60 70 100 70]{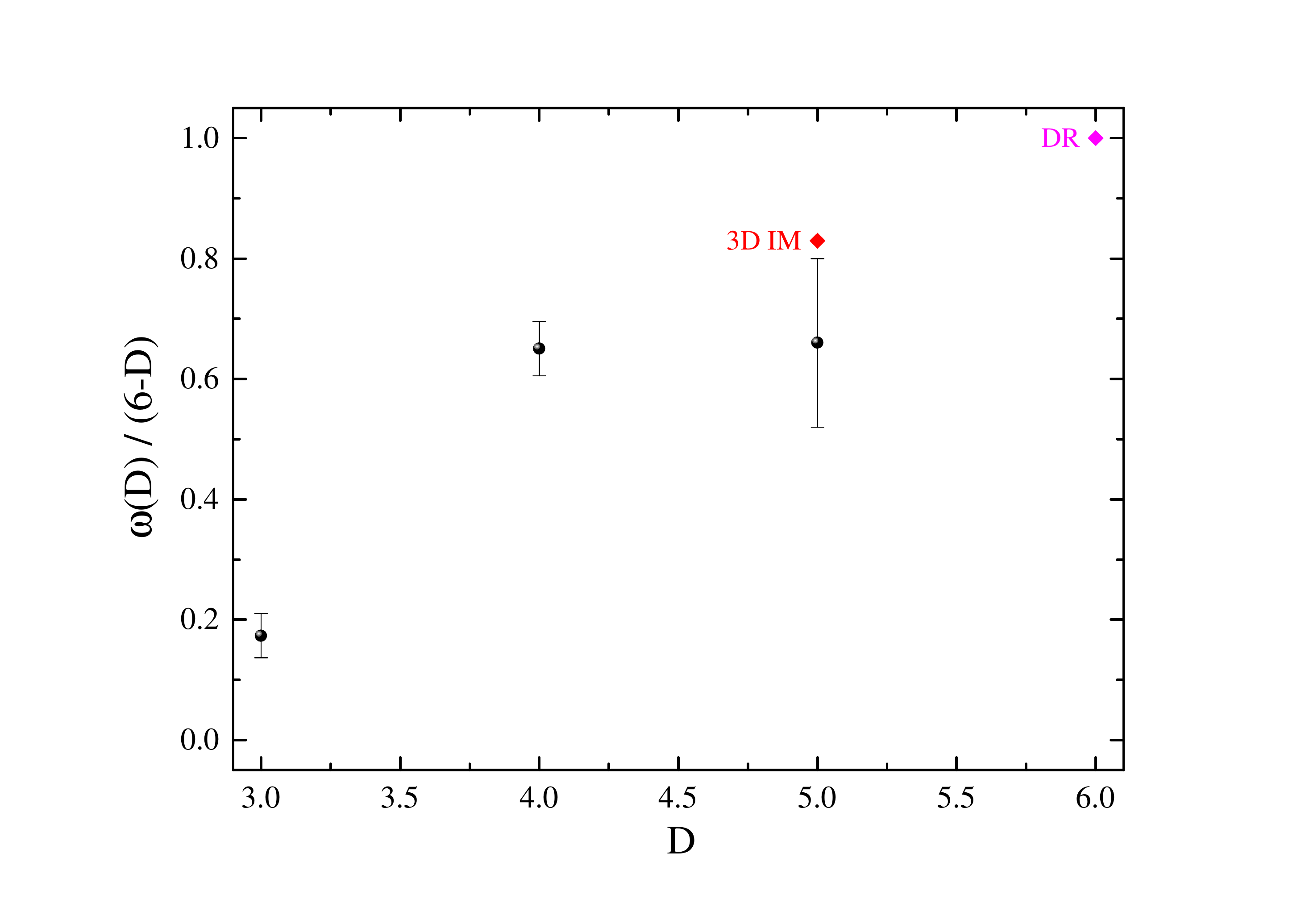}
\caption{\label{fig:omega-RFIM} The corrections to scaling exponent $\omega$,
  as computed from the RFIM in $D=3$~\cite{fytas:13}, $D=4$~\cite{fytas:16}
  and $D=5$~\cite{fytas:17} in units of $6-D$ versus the space dimension. If
  we explicitly assume dimensional reduction (DR), we also have an exceedingly more
  accurate result for $D=5$ (from the three-dimensional pure Ising
  model (3D IM)~\cite{kos:16}) and an exact result at $D=6$.}
\end{figure}

As discussed in the conclusions of the main part, dimensional reduction
suggests that $\omega(D)=\epsilon + {\cal O}(\epsilon^2)$, with
$\epsilon=6-D$. Indeed, Fig.~\ref{fig:omega-RFIM} strongly suggests that the
dimensional-reduction prediction is sensible, because $\omega(D)/(D-6)$ seems
a very smooth function of $D$. We do not find any indication for a zero of
$\omega(D)$ near $D=5$. It is our impression that such a zero, which we do not
see, would be a direct prediction of the Scenario 3 discussed in the main
paper.

\section{Exponent $\omega$ for the pure Ising model in $D=2$}
 \label{sect:omega-Ising-2D}

Paradoxically, it is not trivial to determine the scaling corrections exponent
$\omega$ in the $D=2$ pure Ising model, which is one of the best known models
in Statistical Mechanics.

The difficulty lies in that the leading correction to scaling seems to have a
somewhat unusual origin. Consider, for instance, the magnetic susceptibility
$\chi$ as computed at the critical point for a system of linear dimension $L$.
It is expected to scale as
\begin{equation}\label{eqSI:Ising2D-1}
\chi \sim  A L^{2-\eta} + C\,,
\end{equation}
where $\eta=1/4$ is the anomalous dimension, $A$ is a scaling amplitude and
$C$ is a constant term due to the analytic part of the free-energy
density. Eq.~\eqref{eqSI:Ising2D-1} can be cast as well in the typical form for
scaling-corrections studies (see, e.g., Ref.~\cite{amit:05}):
\begin{equation}\label{eqSI:Ising2D-2}
\chi \sim \ L^{2-\eta} (A\ +\ C L^{-\omega})\quad,\quad \omega=2-\eta=7/4.
\end{equation}
However, this exponent $\omega=7/4$ is not related to any irrelevant operator,
but to the analytic part of the free-energy. Hence, the reasoning leading us
to Eq.~\eqref{eqSI:Ising2D-2} makes sense only if the $\omega$ exponents arising
from \emph{all} the irrelevant operators are larger than $7/4$. Only under
this assumption the leading corrections to scaling would be given by
Eq.~\eqref{eqSI:Ising2D-2}.

Now, it is well known that an operator  associated to the dilution
for the q-Potts models in  $D=2$ (the $q=2$ Potts model is the Ising model)
has dimension $10/3$ and then $\omega=-(D-10/3) =4/3$~\cite{nienhuis:82}.
According to the discussion above, the leading corrections to scaling would
then be given by $\omega=4/3$, rather than $7/4$. However, we think this
is not the case, due to a number of theoretical and numerical reasons:
\begin{itemize}
\item This dilution operator is outside of the main Kac table of operators for
  the Ising model.  Thus it is not produced by other operators (susch as the
  Identity, spin or energy operators) and then it is expected that this
  operator does not contribute to the corrections to scaling. Note that, on
  the contrary, the operator is inside the Kac table for other Conformal Field
  Theories (CFT), such as the 3-Potts model~\cite{dotsenko:84}, for instance.
  In fact, in the limit $q\to 4$, the critical points for Potts and the
  tricritical Potts (which corresponds to the dilution fixed point) merge and,
  indeed, the dilution operator has a dimension 2 in this limit.  It is one
  example for which one finds $\omega = 0$.
\item The above analytical reasoning was confirmed in Ref.~\cite{blote:88}.
  which considered (numerically) various extension of the Ising model
  (antiferromagnetic Ising model in a magnetic field and the Blume-Capel
  model). The exponent $\omega=4/3$ was \emph{not} found in any of these
  models (rather, a correction $\omega \simeq 2$ was identified).  Indeed, the
  authors of Ref.~\cite{blote:88} concluded that the dilution contribution to
  the correction to scaling is indeed given by an exponent $\omega=4/3$, but
  with amplitudes proportional to $(q-2)$ and thus is absent for the Ising
  model, in agreement with CFT predictions. This scenario was supported by
  simulations of the random-cluster model for $q$ close to 2.
\item A recent, very-high accuracy simulation~\cite{shao:16} found again $\omega=7/4$.
\end{itemize}

\bibliography{biblio}

\end{document}